\documentclass[12pt,a4paper]{article}
\usepackage{authblk}
\usepackage{indentfirst}
\usepackage{graphicx}
\usepackage{epsfig}

\usepackage{latexsym}
\usepackage{amsmath}
\usepackage{amssymb}
\usepackage{amsfonts}
\usepackage{mathrsfs}
\usepackage{pifont}

\usepackage{amsthm}
\usepackage{float}
\usepackage{subfigure}
\usepackage{color}
\usepackage{multicol}

\usepackage[colorlinks,linkcolor=blue,filecolor=black,citecolor=blue]{hyperref}

\usepackage{ulem}
\title{A Time Delay Dynamical Model for Outbreak of 2019-nCoV and the Parameter Identification}
\author[1]{Yu Chen \thanks{Email: yuchen@sufe.edu.cn}}
\author[2]{Jin Cheng \thanks{Corresponding author: jcheng@fudan.edu.cn}}
\author[1]{Yu Jiang \thanks{Email: jiang.yu@mail.shufe.edu.cn}}
\author[1]{Keji Liu \thanks{Email: liu.keji@sufe.edu.cn; kjliu.ip@gmail.com}}
\affil[2]{{\small School of Mathematics, Shanghai University of Finance and Economics, Shanghai, China}}
\affil[3]{{\small School of Mathematical Sciences, Fudan University, Shanghai, 200433, China}}

\date{}
\begin{document}
\maketitle

\section*{Abstract}
In this paper, we propose a novel dynamical system with time delay to describe the outbreak of 2019-nCoV in China. One typical feature of this epidemic is that it can spread in latent period, which is therefore described by the time delay process in the differential equations. The accumulated numbers of classified populations are employed as variables, which is consistent with the official data and facilitates the parameter identification. The numerical methods for the prediction of outbreak of 2019-nCoV and parameter identification are provided, and the numerical results show that the novel dynamic system can well predict the outbreak trend so far.  Based on the numerical simulations, we suggest that the transmission of individuals should be greatly controlled with high isolation rate by the government.

\paragraph{Key words:} dynamical system, time delay process, parameter identification, outbreak prediction, isolation.

\section{Introduction}
Coronaviruses are enveloped single-stranded positive-sense RNA viruses belonging to the family Coronaviridae and the order Nidovirales which are discovered and characterized in 1965 and are broadly distributed in mammals and birds. In humans, most of the coronaviruses cause mild respiratory infections, but rarer forms such as the ``Severe Acute Respiratory Syndrome'' (SARS) in China and the ``Middle East Respiratory Syndrome'' (MERS) in Saudi Arabia and South Korea had cased more than 10000 cumulative cases in the past two decades.
Although many coronaviruses had been identified and characterized, they might be a tip of the iceberg and lots of potential severe and novel zoonotic coronaviruses needed to be revealed.

In December 2019, a series of pneumonia cases of unknown cause  emerged in Wuhan, the capital of Hubei province and one of largest cities in the central part of China. The Wuhan Municipal Health Commission reported 27 cases of viral pneumonia including 7 severely ill cases on December 12th 2019, and the outbreak of pneumonia began to attract considerable attention in the world. And the causative agent identified by the Chinese authorities was designated 2019 novel coronavirus (2019-nCoV) by the World Heath Organization (WHO) on January 10th 2020.

On January 20th 2020, the Chinese government has revised the law provisions of infectious diseases to add the novel 2019-nCoV as class A agent. And a series of non-pharmaceutical interventions were implemented, namely, isolation of symptomatic person, strictly prohibit the travel in the Hubei province, partially shut down the public transport in many cities, etc. However, the effectiveness and efficiency of these interventions during the early stage is questionable.
So far, there are more than 3000 confirmed cases in Wuhan and more than 11000 confirmed cases in China, and several exported cases have been confirmed in many other countries including Japan, South Korea, Singapore, USA, Canada, Germany, France, UK, Spain, etc. 

Because the prediction for tendency of 2019-nCoV is of great importance at present, we state the three distinct features in our novel dynamic model as follows,

\begin{enumerate}
	\item One distinct feature of this epidemic is that it can spread during the latent period, which is different from SARS in 2003. Therefore, many classical models such as SIR\cite{Bere1995, canto2017, ma2004}, SEIR\cite{ma2004} and SEIJR\cite{Ding2004} are not appropriate to describe the outbreak of 2019-nCoV in China. Thanks to the reliable data of average latent period and average time of treatment are available, we can apply the \textit{time delay process} in our novel dynamical system to describe the typical feature appropriately.
	\item The National Health Commission of China only report the accumulated numbers such as diagnosed cases, cured cases and died cases. In addition, the accumulated numbers are more stable in numerical simulations than the instantaneous increment data. We thus employ the \textit{accumulated numbers in time as variables} in our novel dynamic model.
	\item Chinese government has implemented strong provisions of isolation for suspected cases and contacted individuals. Accordingly, the \textit{isolation process} is introduced in our novel model to measure this phenomenon. 
\end{enumerate}

The remaining part of this paper is organized as follows. In Section \ref{sec::model}, we propose the novel dynamic model of the outbreak of 2019-nCoV in China. The novel approach for estimating the future number of diagnosed people with this model is provided in Section \ref{sec::imp}. Based on the official data, several numerical experiments are exhibited in Section \ref{sec::imp} to verify the effectiveness and accuracy of our estimation scheme. 
Finally, we present some concluding remarks in section \ref{sec::discussion}.

\section{A Novel Dynamical System with Time Delay}\label{sec::model}
In this section, a novel dynamic system with time delay to describe the outbreak of 2019-nCoV would be presented in details.
We first state our dynamical model by providing following data description, denotations and assumptions.

\paragraph{Data Description:\\[2mm]}

The data employed in our model are derived from the National Health Commission of China and the Health Commission of each province and city in China. Moreover, the data is from Jan. 23rd 2020 to Feb. 9th 2020 and including the cumulative diagnosed people, the cumulative cured people and the cumulative dead people.

\paragraph{Denotations:} 

\begin{itemize}
	\item $I(t)$: accumulated number of infected people at time $t$;
	\item $J(t)$: accumulated number of diagnosed people at time $t$;
	\item $G(t)$: currently isolated people who are infected but still in latent period at time $t$;
	\item $R(t)$: accumulated number of cured people at time $t$.
\end{itemize}

\paragraph{Assumptions:}
\begin{enumerate}
	\item We assume the infected person can transmit the coronavirus to others at a spread rate $\beta$, which is defined by the average amount of people becoming infected by this person in unit time. 
	
	\item The infected people averagely experience a latent period of $\tau_1$ days before they have significant symptoms. It is assumed that upon the appearance symptoms, the infected person will seek for treatment and therefore become diagnosed people.
	
	\item Some of the infected people would be exposed in the latent period $\tau_1$ until they are diagnosed. The average exposed period of these people are $\tau_1-\tau_1'$ days, which means they would be diagnosed in the next $\tau_1'$ days. Some other part of the infected people are isolated during latent period according to investigation of diagnosed cases. 
	
	\item The accumulated diagnosed people $J(t)$,  no matter they are isolated before diagnosed or not, are consist of the population infected at time $t-\tau_1$ (averagely).
	
	\item We assume that a person once isolated or in treatment, the individual would no longer transmit the coronavirus to others. Consequently, the exposed people at time $t$ are $I(t)-G(t)-J(t)$.
	
	\item It is $\tau_2$ days in average for the diagnosed people become cured with rate $\kappa$ or dead with rate $1-\kappa$.
	
\end{enumerate}

\begin{figure}[H] 
	\begin{center}
		\includegraphics[width=0.8\textwidth]{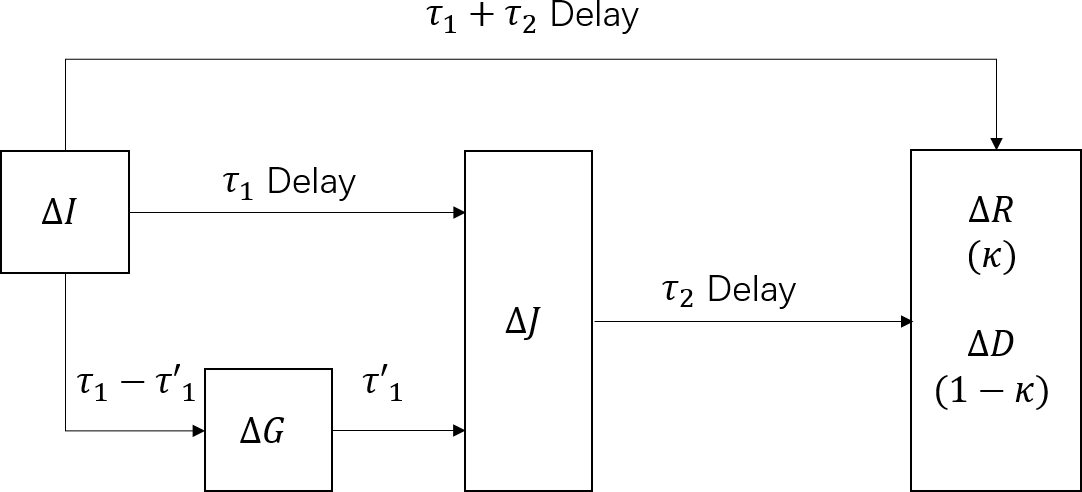}
	\end{center}
	\caption{The demonstration of the novel dynamic system.}
	\label{illustration} 
\end{figure}

Based on the notations and assumptions, the novel dynamic system with time delay to describe the outbreak of 2019-nCoV is demonstrated in Figure \ref{illustration}. 

\paragraph{Dynamic Model of Closed System without External Source:\\[2mm]}

Suppose the dynamics of these populations are described by the following ODE system,
\[
\frac{\mathrm{d}\textbf{Y}}{\mathrm{d}t}=\textbf{F}(\textbf{Y},\tau_1,t,\tau_1,\tau_1', \tau_2),
\]
where $\textbf{Y}=[I,J,G,R]$ is the variable vector and the vector of right-hand side term $\textbf{F}=[I_1,J_1,G_1,R_1]$ represents the instant change of each variable. The detailed explanations for each term of $\textbf{F}$ are stated in the following paragraphs.

At time $t$, the exposed people that may transmit to others are $I(t)-G(t)-J(t)$. According to the definition of spread rate $\beta$ (see Assumption 1), the instant increment of infected people at time $t$ is 
\[
I_1(t)=\beta \Big(I(t)-G(t)-J(t)\Big),
\]
where the spread rate $\beta$ is a fixed constant. In a general case, $\beta$ might depend on time $t$ since viruses activities and environment may change. 

No matter having been isolated or not, the cumulative diagnosed people $J(t)$ all come from the previously infected population. Therefore, the increment of $I(t)$ at time $t'$ ($t'<t$), i.e., $I_1(t')$, would contribute to $J_1(t)$, which means $J_1(t)$ depends on the history of $I_1(t)$. If the average delay between infected time and diagnosed time is $\tau_1$, the specific form of $J_1$ can be denoted as 
\begin{equation}
J_1(t)=\gamma\int_0^t h_1(t-\tau_1,t')I_1(t')\mathrm{d}t',\label{1delay}
\end{equation}
where $\gamma$ is the morbidity, and $h_1(\hat{t},t')$ ($\hat{t}=t-\tau_1$) is a distribution which should be normalized as 
\[\int_0^t h_1(\hat{t}, t')\mathrm{d}t'=1,\quad \hat{t}\in (0,t).\]
We are able to observe that $h_1(\hat{t},t)$ can be regarded as the probability distribution of infection time $t'$, and we usually take the normal distribution $h_1(\hat{t},t')=c_1 e^{-c_2 (\hat{t}-t')^2}$ with $c_1$ and $c_2$ be constants. In the simplest case, $h_1$ can also be the $\delta$-function $h_1(\hat{t},t')=\delta(\hat{t}-t')$, which means that every infected individual experienced the same latent period and treatment period. 

The instant change in $G(t)$ is defined in the following form
\begin{equation}
G_1(t):=\ell I_1(t)-\int_0^t h_2(t-\tau'_1,t')G(t') \mathrm{d}t',\label{2delay}
\end{equation}
where $\ell$ is the rate of isolation for the currently exposed people, and $h_2(\hat{t},t')=c_3 e^{-c_4 (\hat{t}-t')^2}$ with $c_3$ and $c_4$ be constants. This means some of the exposed infectors are newly isolated, and some existent isolated infectors are diagnosed and sent to hospital for treatment.  The time delay term $\int_0^t h_2(t-\tau'_1,t')G(t') \mathrm{d}t'$ stands for the newly diagnosed people among $G(t)$ depending on the history of $G(t)$. 


As illustrated above, the accumulated cured people at time $t$ comes from the ones infected at $t-\tau_1-\tau_2$ (in average). We apply the time delay term 
\begin{equation}
\kappa\int_0^t h_3(t-\tau_1-\tau_2,t')I_1(t')\mathrm{d}t'\label{3delay}
\end{equation}
to describe $R_1$, where $h_3(\hat{t},t')=c_5 e^{-c_6 (\hat{t}-t')^2}$ with $c_5$ and $c_6$ be constants.

Substitute $I_1$ into the time delay terms in \eqref{1delay},\eqref{3delay}, we arrive at the following novel dynamic closed system without external source
\begin{equation}
\left\{
\begin{aligned}
\frac{\mathrm{d}I}{\mathrm{d}t}&=\beta\Big(I(t)-J(t)-G(t)\Big),\label{eq::modelstart}\\[2mm]
\frac{\mathrm{d}J}{\mathrm{d}t}&=\gamma \int_0^t h_1(t-\tau_1,t')\beta\Big(I(t')-J(t')-G(t')\Big)\mathrm{d}t',\\[2mm]
\frac{\mathrm{d}G}{\mathrm{d}t}&=\ell \Big(I(t)-J(t)-G(t)\Big)-\int_0^t h_2(t-\tau'_1,t')G(t') \mathrm{d}t',\\[2mm]
\frac{\mathrm{d}R}{\mathrm{d}t}&=\kappa\int_0^t h_3(t-\tau_1-\tau_2,t')\beta\Big(I(t')-J(t')-G(t')\Big)\mathrm{d}t'.
\end{aligned}
\right.
\end{equation}

We can observe that the evolution equations for $I(t)$, $J(t)$ and $G(t)$ are coupled in the system \eqref{eq::modelstart}. 
Although $R(t)$ is not involved in the equations for $I(t)$, $J(t)$ and $G(t)$, the evolution of $R(t)$ employs $I(t)$, $J(t)$ and $G(t)$ as input. 
In other words, $R(t)$ have no impacts on the other variables in the forward problem, and the observation of $R(t)$ may help inferring the parameters and initial conditions in the inverse problem. 
It is noted from the official data, the cumulative diagnosed people $J(t)$ and the cumulative cured people $R(t)$ are always available, while $I(t)$ and $G(t)$ are usually unobtainable since they are hard to measure. In practical applications, we thus apply the cumulative diagnosed people $J(t)$ and the cumulative cured people $R(t)$ in the parameter identification.

\section{The Reconstruction and Prediction Scheme}\label{sec::imp}

In this section, we shall present the methods to reconstruct parameters and predict the tendency of outbreak for the 2019-nCoV.
Suppose we know the proper parameters $\{\beta,\gamma,\kappa,\ell,\tau_1,\tau'_1,\tau_2\}$ and the  initial conditions $\{I(t_0), G(t_0), J(t_0),R(t_0)\}$ of the novel dynamic system \eqref{eq::modelstart}, the cumulative diagnosed people $J(T)$ and cumulative cured people $R(T)$ at any given time $T$ are readily to obtain by solving the novel dynamic system numerically. Moreover, we suggest to implemented the Matlab$\circledR$ inner-embedded program {\bf dde23} to solve the novel dynamic system.

In practice, it's reasonable to have following assumptions for the initial conditions and the parameters:
\begin{enumerate}
	
	\item{\bf initial conditions:} $I(t_0)=5$, $G(t_0)=J(t_0)=R(t_0)=0$. On the initial day $t_0$, we suppose $5$ people are infected the 2019-nCoV from unknown sources. Moreover, the diagnosed, isolated and recovered people is 0 on the initial day. In the numerical simulation, we further assume that there are no isolation meausres implemented before $T=t_0+15$. 
	\item {\bf parameters:} 
	According to the present data, the morbidity is relatively high with $\gamma=0.99$. The average latent period $\tau_1$ and treatment period $\tau_2$ are also regarded as known according to the official data. The average period between getting isolated and diagnosed $\tau_1'$ satisfies $0<\tau_1'<\tau_1$. The known parameter set is summarized in Table \ref{par-known}. 
	\begin{table}[H] 
		\begin{center} 
			\begin{tabular}{cccc}  
				\hline  
				$\gamma$ &$\tau_1$ &$\tau'_1$ & $\tau_2$  \\  
				\hline  
				$0.99$ & $7$ &$4$ & $12$  \\  
				\hline  
			\end{tabular} 
		\end{center}
		\caption{The values for known parameters.}\label{par-known}
	\end{table}
\end{enumerate}

Therefore, the set of parameter that need to be reconstructed is reduced to  
$$\theta:=[\beta,\ell]\quad \text{and}\quad \kappa,$$
and our parameter identification problems come to following two optimization problems,
\begin{equation}\label{eq::op1}
\min_{\theta}\|J(\theta;t)-J_{Obs.}\|_2,
\end{equation}
and
\begin{equation}\label{eq::op2}
\min_{\kappa}\|R(\theta,\kappa;t)-R_{Obs.}\|_2.
\end{equation}
Here, $J_{Obs.}$ and $R_{Obs.}$ are the daily data reported by the National Health Commission of  China. 
Now we are ready to state our reconstruction and prediction scheme as follows\\[2mm]
{{\bf Reconstruction and Prediction Scheme}:} 
\begin{itemize}
	\item[Step 1.]  Based on the official data $J_{Obs.}$, we solve the optimization problem \eqref{eq::op1} to acquire the reconstructed parameter $\theta^*$.
	
	\item[Step 2.] Based on the official data $R_{Obs.}$ and the recovered $\theta^*$, the optimization problem \eqref{eq::op2} is solved to obtain the estimated parameters $\kappa^*$.
	
	\item[Step 3.] With the reconstructed $[\theta^*,\kappa^*]^{\rm T}$ , one can attain the prediction of $\{J(t),R(t)\}$ and also $\{I(t), G(t)\}$ by solving the system \eqref{eq::modelstart} numerically.
\end{itemize}

It is worth mentioning that, either matter the Levenberg-Marquad (L-M) method or the Markov chain Monte Carlo (MCMC) method would work well for solving the optimization problems \eqref{eq::op1} and \eqref{eq::op2}. Since those algorithms are quite well known and easy to implement by Matlab$\circledR$, we omit the details which can be found in, for example, \cite{KNO} and \cite{KS}.

\section{Numerical Test}\label{sec::test}
In this section, some numerical experiments are exhibited to verify the effectiveness and accuracy of novel dynamic system.

\paragraph{Basic simulation}

\begin{figure}[H] 
	\begin{center}
		\subfigure[With isolation]
		{
			\includegraphics[width=0.45\textwidth]{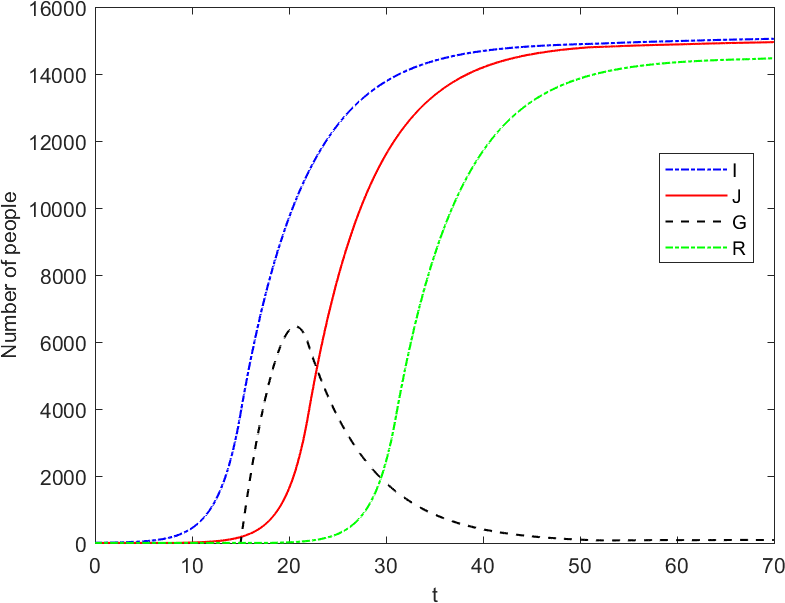}
			\label{ideal-1} 
		}
		\subfigure[Without isolation]
		{
			\includegraphics[width=0.43\textwidth]{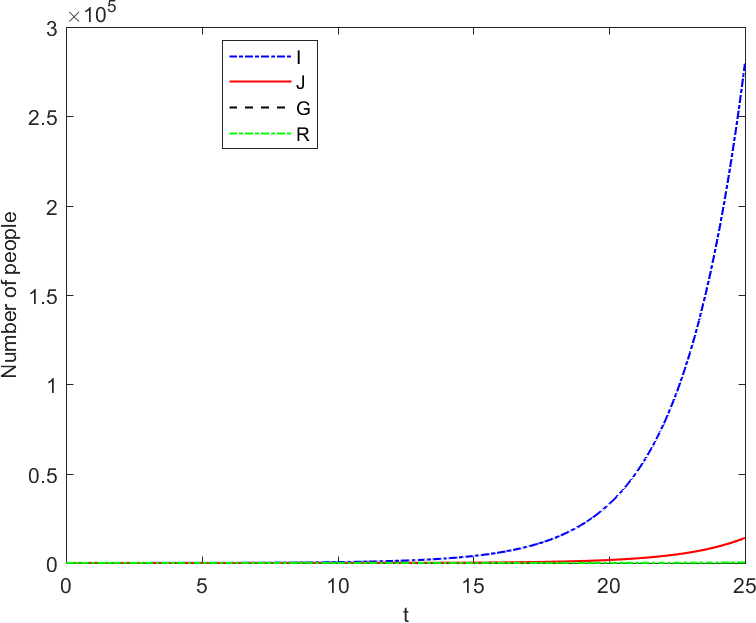}
			\label{ideal-2} 
		}
	\end{center}
	\caption{Model simulation}
	\label{Ideal} 
\end{figure}
For given $\beta=0.45$, $t_0=0$ and the parameters in Table \ref{par-known}, Figure \ref{Ideal} shows the simulation with different rate of isolation. The case with isolation ($\ell=0.6$) is exhibited in Figure \ref{ideal-1}. It can be observed that in this case the infected people would increase sharply at first and mildly after $t=20$ days. The cumulative infected people would finally tend to about 14000 after $t=40$ days. The evolutions of cumulative diagnosed and cured people possess similar trends with the corresponding lag in time. However, in the case without isolation, the accumulated infected people would tend to infinity, see Figure \ref{ideal-2}.

\begin{figure}[H] 
	\begin{center}
		\subfigure[Mainland of China]
		{
			\includegraphics[width=0.4\textwidth]{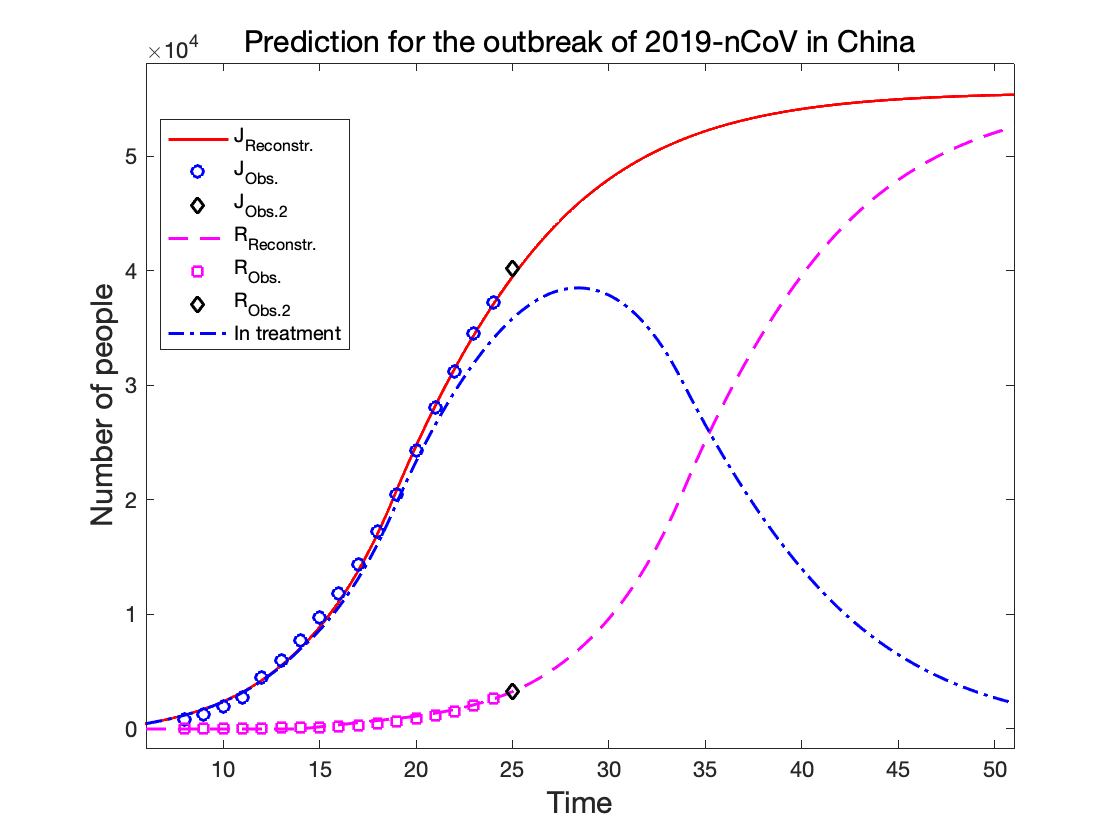}\label{fig::china}
		}
		\subfigure[City 1]
		{
			\includegraphics[width=0.4\textwidth]{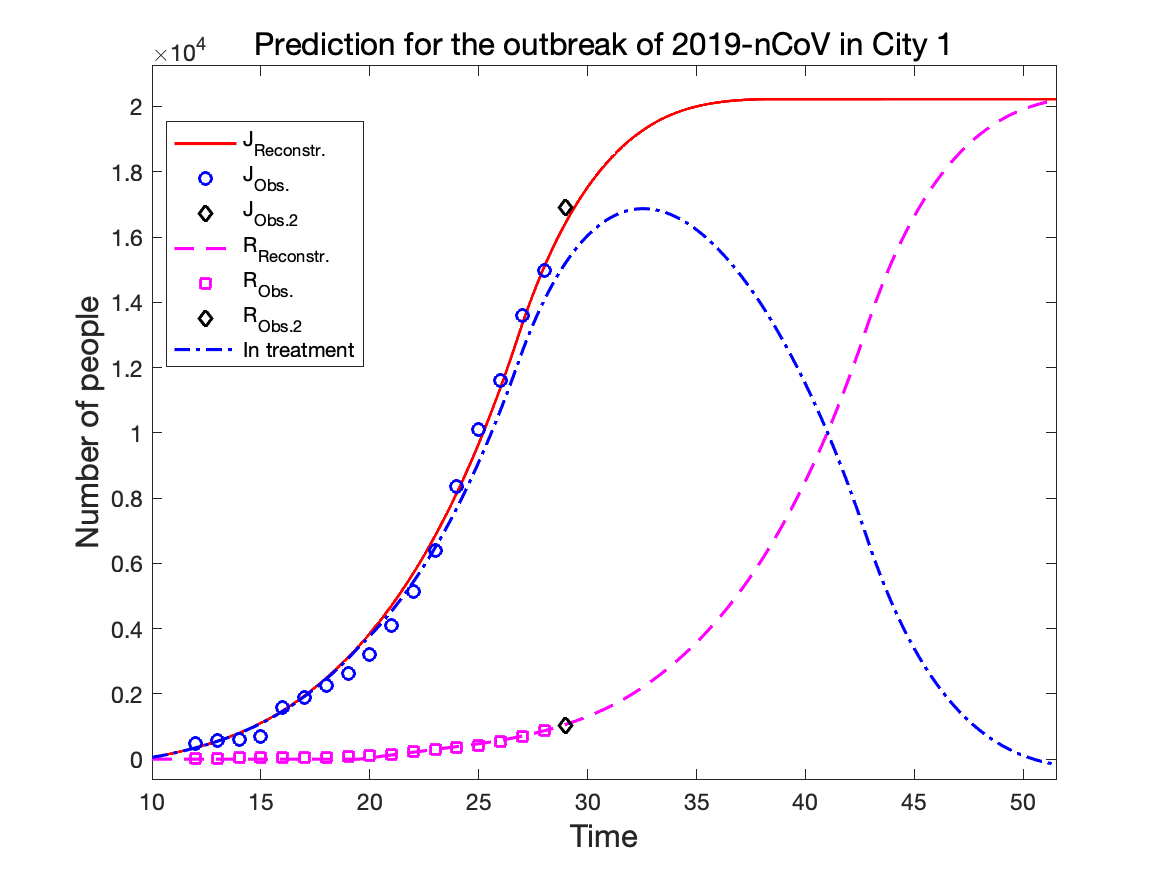}\label{fig::city1}
		}\\
		\subfigure[City 2]
		{
			\includegraphics[width=0.4\textwidth]{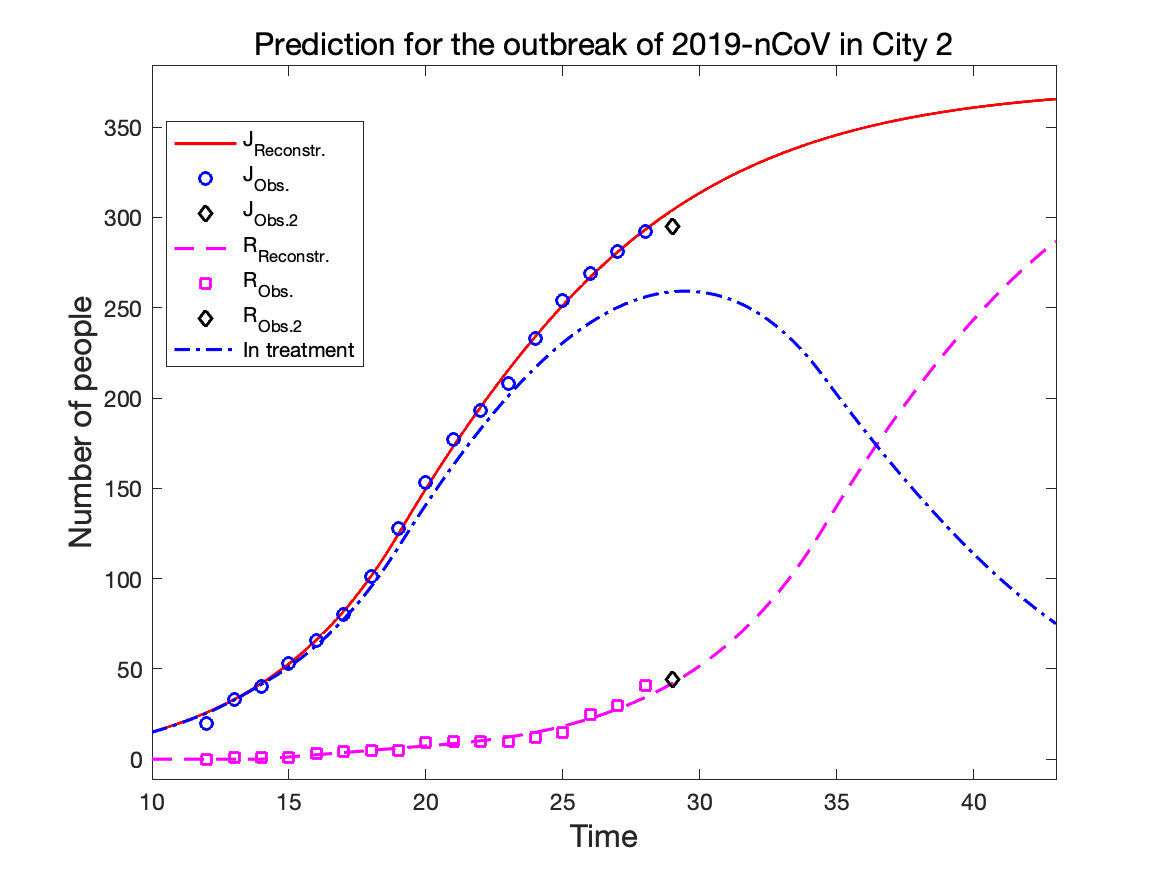}\label{fig::city2}
		}
		\subfigure[Province 1]
		{
			\includegraphics[width=0.4\textwidth]{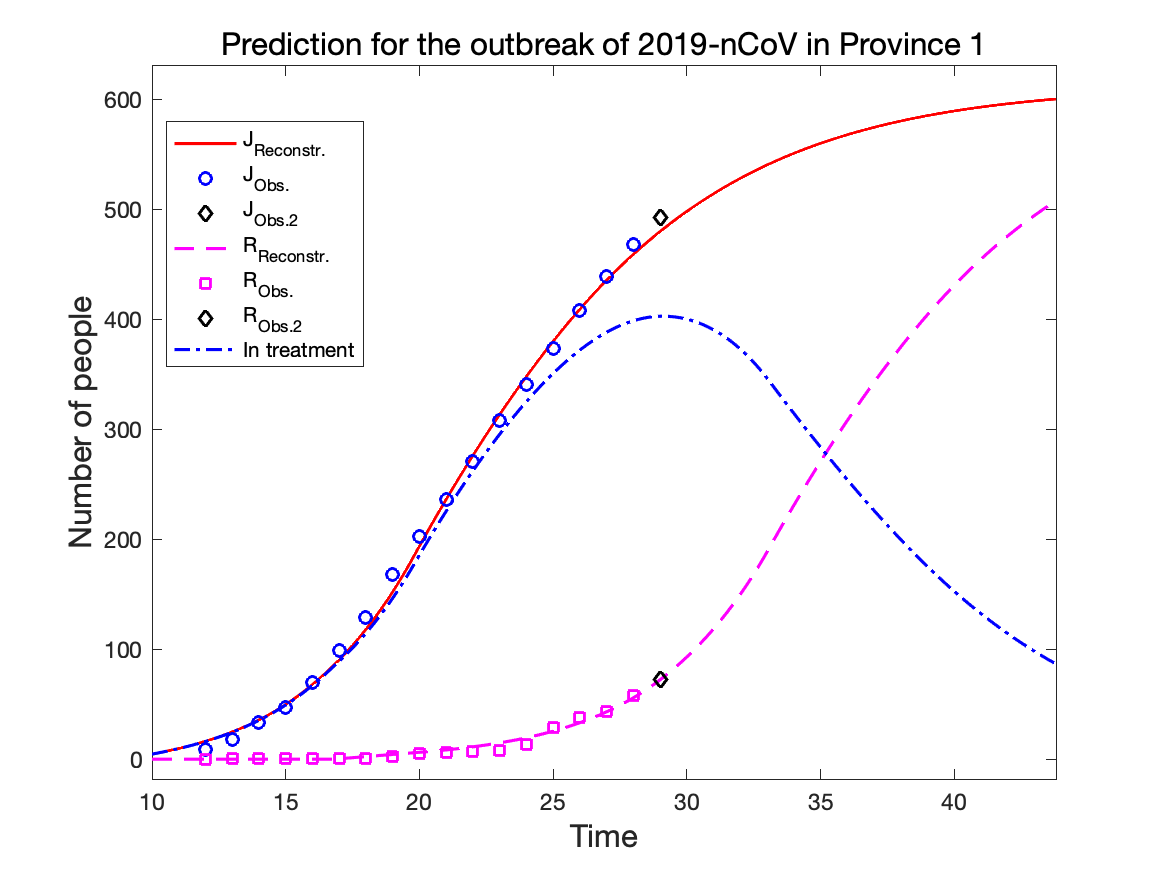}\label{fig::province1}
		}
	\end{center}
	\caption{Reconstruction and Prediction. $J_{Obs.}$ and $R_{Obs.}$ are the data of cumulative diagnosed and cured people for reconstructions. $J_{Obs.2}$ and $R_{Obs.2}$ are displayed for evaluations of the predictions. $J_{Reconstr.}$ and $R_{Reconstr.}$ correspond to the estimated values of diagnosed and cured cases.}\label{real-national} 
\end{figure}

\paragraph{Global outbreak} 
In this example, we provide some numerical results with the estimated parameters and the official data in this event.  We apply the national data covering 10 days to infer those parameters which are presented in Table \ref{tab::parameter}. The case of 18th day is demonstrated in Figure \ref{fig::china} to test the accuracy of prediction.

\paragraph{Local outbreak} In this example, we provide some numerical results under reconstructed parameters with observation of official data of several local city/provinces. City 1 is the source of this event and has the most serious situation so far. City 2 is the most developed city in China, while province 1 is close to city 2.
We also employ the data covering 17 days to infer those parameters which are presented in Table \ref{tab::parameter}. 
The cases of 18th day are separately demonstrated in Figure \ref{fig::city1}-\ref{fig::province1} to illustrate the accuracy of predictions.

Look through the Figure \ref{real-national} and reconstructed parameters in Table. \ref{tab::parameter}, we can conclude as follows
\begin{enumerate}
	\item by the newly proposed time delay dynamic system, it is possible to predict the tendency of outbreak for the 2019-nCoV. And the numerical results reveal the situation would be better and better.
	\item the estimated parameters show the fact that the evaluation of event is highly dependent on the population, size and public policy carrying on by the local governments.
\end{enumerate}
\begin{table}[H] 
	\begin{center} 
		\begin{tabular}{cccc}  
			\hline  
			Region & $\beta$ & $\ell$ &$\kappa$  \\  
			\hline  
			Mainland of China &  0.2274  &  0.3052 & 0.9692 \\
			City 1 (Wuhan)  &  0.2079   &  0.4236  & 0.9697\\
			City 2 (Shanghai)  &  0.2529  &  0.3042  &0.9703\\
			Province 1 (Jiangsu)  &  0.2527 &  0.3048 &0.9701  \\
			\hline  
		\end{tabular} 
	\end{center}
	\caption{Estimated parameters.} 
	\label{tab::parameter} 
\end{table}

\section{Concluding Remarks}\label{sec::discussion}

We propose a novel dynamic system with time delay to predict the trend of outbreak for the 2019-nCoV. In this model, the instantaneous increment of cumulative diagnosed people depends on the history of cumulative infected people, by which the latent period can be taken into consideration. 
The numerical simulation and parameter identification were carried out to verify the effectiveness and accuracy of the novel dynamic system. In addition, this model can well approximate the true data in this event, which may further predict the tendency of event. 

The present result can also provide some suggestions for the prevention and control of the 2019-nCoV. Firstly, the isolation is essential in controlling the spread of 2019-nCoV. With the isolation rate 0.5, it is possible to end the outbreak of 2019-nCoV in around 50 Days. Secondly, the needs of medical resources would arrive at the peak at about 35th day from the beginning of outbreak. 
In the case of transmission within latent period, the time delay term is effective in modeling the process. 
At present, the parameters involved in the novel system are time-independent. In the future work, the change in isolation strength and spread rate may be assumed to depend on time so as to improve the agreement between real data and estimated solution. Besides, in this novel system, $J(t)-R(t)$ is the amount of people in the treatment, which is also possible to measure and can be employed in the inverse problem. Furthermore, some external source terms may be added into the model to reflect the uncertainties of the process.

\section*{Acknowledgments}
This work was supported in part by the National Science Foundation of China  (NSFC: No. 11971121) and the Science and Technology Commission of Shanghai Municipality under the ``Shanghai Rising-Star Program'' No. 19QA1403400. The authors thank for the helpful discussions with Dr. Yue Yan, Dr. Boxi Xu, Ms. Xinyue Luo and Mrs. Jingyun Bian in Shanghai University of Finance and Economics. Cheer up Wuhan!

\addcontentsline{toc}{section}{

\end{document}